# ATDD: Fine-Grained Assured Time-Sensitive Data Deletion Scheme in Cloud Storage

Zhengyu Yue, Yuanzhi Yao*, Weihai Li, Nenghai Yu
School of Cyber Science and Technology,
University of Science and Technology of China, Hefei 230027, China
e-mail: yzyaibb@mail.ustc.edu.cn, {yaoyz, whli, ynh}@ustc.edu.cn

***Abstract*—With the rapid development of general cloud services, more and more individuals or collectives use cloud platforms to store data. Assured data deletion deserves investigation in cloud storage. In time-sensitive data storage scenarios, it is necessary for cloud platforms to automatically destroy data after the data owner-specified expiration time. Therefore, assured time-sensitive data deletion should be sought. In this paper, a fine-grained assured time-sensitive data deletion (ATDD) scheme in cloud storage is proposed by embedding the time trapdoor in Ciphertext-Policy Attribute-Based Encryption (CP-ABE). Time-sensitive data is self-destructed after the data owner-specified expiration time so that the authorized users cannot get access to the related data. In addition, a credential is returned to the data owner for data deletion verification. This proposed scheme provides solutions for fine-grained access control and verifiable data self-destruction. Detailed security and performance analysis demonstrate the security and the practicability of the proposed scheme.**

## I. Introduction

With the rapid development of general cloud services, cloud storage services have significant advantages in terms of convenient data sharing and cost reduction [1, 2]. However, this new form of data storage brings new challenges to data confidentiality protection. As customers (individuals or entities) are separated from cloud data and cloud servers, they lose direct control over their data [3]. A large number of time-sensitive data, such as health data, financial data, trade secrets or privileged communications have been outsourced to cloud storage, and data owners cannot fully trust the access control provided by cloud storage. This leads to many data privacy and other security issues [4–6]. To solve these problems, a comprehensive solution should be designed to provide fine-grained access control for the duration of user defined authorization and assuredly delete expired time-sensitive data timely.

Perlman [7] designed the first guaranteed data deletion file system, in which data was created using an expiration time. Tang *et al.* [8] proposed a policy-based assured data deletion scheme FADE. The downside of FADE is that it fails to provide fine-grained access to files. Mo *et al.* [9] proposed a fine-grained guaranteed deletion scheme, in which a key modulation function using anti-collision hash function is designed. Ateniese *et al.* [10] pointed out that device authentication can use spatial proof to confirm that remote embedded devices have been successfully cleared. However, this method is only for embedded devices with limited memory and cannot be directly applied to cloud storage environment. Yu *et al.* [11] proposed an attribute-based fine-grained cloud data deletion scheme. However, the deletion verification process is equivalent to letting the data owner run the attribute-based decryption process completely, which is not very friendly to the data owner and the scheme cannot delete data periodically. None of the above schemes can achieve fine-grained assured time-sensitive data deletion.

Inspired by the Timed Release Encryption (TRE) [12, 13], our proposed fine-grained assured time-sensitive data deletion (ATDD) scheme inserts a time trapdoor in the root node of the Ciphertext-Policy Attribute-Based Encryption (CP-ABE) access policy. In [12], the trapdoor embedded in logic gates can be removed, and the access privilege transfer is only determined by the attribute set. In our work, at the time point set by the data owner, the cloud service center updates the time trapdoor parameters of the access policy in time so that the data cannot be decrypted by any user. Thus, the purpose of destroying data can be achieved. In addition, after the cloud data center destroys the data, it returns the deletion certificate constructed by the Merkle Hash Tree (MHT) [14] to the data owner. After receiving the certificate, the data owner can confirm that the cloud data center has indeed deleted the data through simple calculation. The proposed scheme ATDD meets the design goal of assured deletion of expired time-sensitive data timely.

The main contributions of this paper can be summarized as follows:

- We propose the scheme ATDD for timed cloud data destruction based on the CP-ABE, which is used to achieve fine-grained access control and effectively delete data at the end of the data life cycle set by the data owner.
- We generate deletion proof by building the MHT for ciphertext. The data owner verifies whether the data has been deleted by using the MHT returned from the cloud server.
- We give detailed security proof and performance analysis of ATDD.

The rest of this paper is organized as follows. Section

II describes the main preliminaries and definitions used in our proposed scheme ATDD. According to the system model and the security model in Section III, the detailed design of ATDD is elaborated in Section IV. The security and the performance of ATDD are analyzed in Section V and Section VI respectively. Finally, we conclude this paper in Section VII.

## II. Preliminaries and Definitions

In this section, we first give a brief review of background information on bilinear maps and the security assumptions defined on it. Then we briefly introduce the Merkle Hash Tree (MHT).

### A. Bilinear Pairings and Complexity Assumption

Our proposed scheme ATDD is built on prime order bilinear groups [15]. Let $\mathbb{G}$ and $\mathbb{G}_T$ be two cyclic groups with prime order $p$. A bilinear map on $\mathbb{G}$ and $\mathbb{G}_T$ is a map $e$: $\mathbb{G}\times\mathbb{G} \rightarrow \mathbb{G}_T$ that holds the following properties:

1) *Bilinearity*: For any $a, b \in \mathbb{Z}_p$ and $g_1, g_2 \in \mathbb{G}$, we have $e\left(g_1^a, g_2^b\right) = e\left(g_1, g_2\right)^{ab}$.
2) *Computability*: There exists a polynomial time algorithm to compute $e\left(g_1, g_2\right)^{ab}$, for any given $g_1, g_2 \in \mathbb{G}$ and $a, b \in \mathbb{Z}_p$.
3) *Non-degeneracy*: If $g$ is a generator of $\mathbb{G}$ then $e\left(g, g\right)$ is also a generator of $\mathbb{G}_T$.

*Definition 1:* Decisional Bilinear Diffie-Hellman (DBDH) assumption.

For a bilinear group of prime order $p$, the DBDH says that no probabilistic polynomial-time algorithm $\mathcal{A}$ can distinguish the tuple $\left(g^a, g^b, g^c, e(g,g)^{abc}\right)$ from the tuple $\left(g^a, g^b, g^c, e(g,g)^{z}\right)$ where $z$ is a random value from $\mathbb{Z}_p$ with more than a negligible advantage. The advantage of $\mathcal{A}$ is

$$\begin{aligned} Adv_{\mathcal{A}} = &\left|\Pr\left[\mathcal{A}\left(g, g^a, g^b, g^c, e(g,g)^{abc}\right) = 0\right]\right. \\ &\left. - \Pr\left[\mathcal{A}\left(g, g^a, g^b, g^c, e(g,g)^{z}\right) = 0\right]\right| \end{aligned}$$

where the probability is over the random choice of the generator $g$ in $\mathbb{G}$ and the random choices of $a$, $b$, $c$, $z$ in $\mathbb{Z}_p$.

### B. Merkle Hash Tree (MHT)

MHT [14] is a tree that stores the hash value of data. The MHT is commonly used to verify the integrity of data blocks. During validation, the root node can be computed by providing all of the siblings on the path to the root node of a given data block. The data block is complete if the calculated root node is equal to the real root node, otherwise the data block is incomplete.

## III. System Model and Security Model

In this section, we present the system model and security model of the proposed scheme ATDD.

### A. System Model Definition

The ATDD system model consists of several entities: some data owners (DO), many data users (DU), a cloud storage server (CS), and a trusted authority (TA) as shown in Fig. 1.

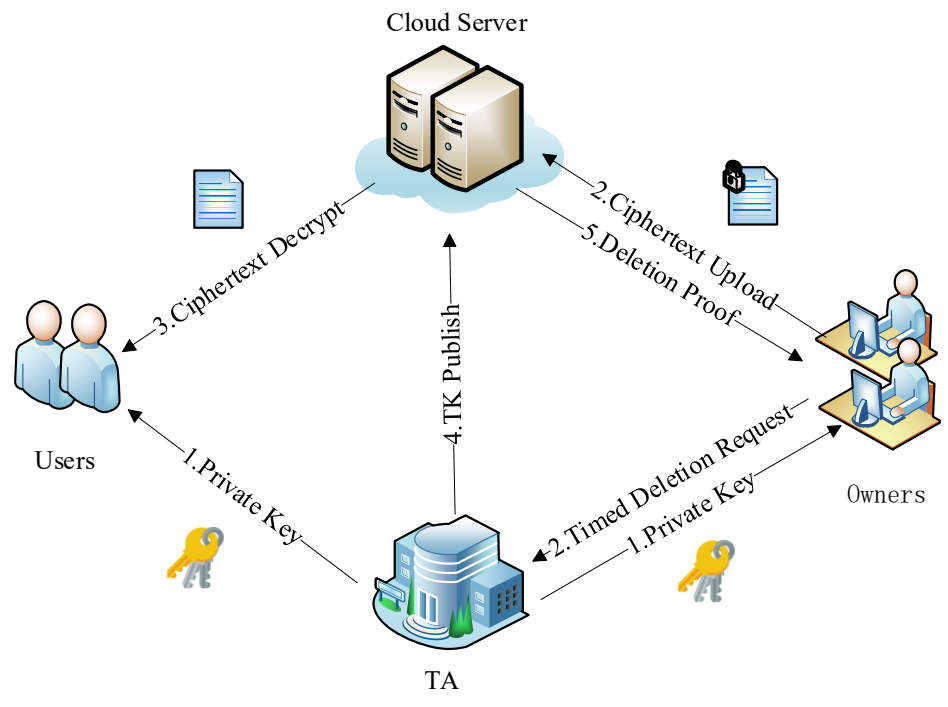

Fig. 1. ATDD system architecture and operations.

### B. Security Model

A formal proof that ATDD can resist chosen plaintext attacks (CPA) can be described as the following security game:

**Initialization**. The challenger $\mathcal{C}$ selects system initialization parameters and starts the DBDH game.

**Setup**. The challenger $\mathcal{C}$ enters system initialization parameters to run the **Setup** algorithm of ATDD. Then, $\mathcal{C}$ gives the public key $PK$ to the adversary $\mathcal{A}$ and keeps the master secret key $MSK$ in secret.

**Phase 1**. The adversary $\mathcal{A}$ submits access structure $\mathbb{A}$ and a set of attributes $S$ which does not satisfy $\mathbb{A}$. The challenger $\mathcal{C}$ generates the security key associated with $S$ and gives it to the adversary $A$.

**Challenge**. The adversary $\mathcal{A}$ submits two equal-length messages $M_0$ and $M_1$. The challenger $\mathcal{C}$ flips a random coin $d \in \{0, 1\}$, and encrypts $M_d$ with $\mathbb{A}$ to output the ciphertext $CT$ which is sent to the adversary $\mathcal{A}$ later.

**Phase 2**. Repeat **Phase 1**.

**Guess**. The adversary $\mathcal{A}$ outputs a guess $b' \in \{0, 1\}$. If $b' = b$, the adversary $\mathcal{A}$ wins the game. The advantage of $\mathcal{A}$ is defined as

$$Adv_{\mathcal{A}} = |\Pr\left[b' = b\right] - \frac{1}{2}|.$$

*Definition 2:* If all polynomial time adversaries have at most a negligible advantage in the above security game, we can conclude that our proposed scheme ATDD is secure.

## IV. Our Proposed Scheme

In this section, we first introduce the implementation structure of ATDD with time trapdoor access policy, and then introduce the specific steps of ATDD algorithm. Table I shows the notations used in the paper.

### A. Access Policy Structure

Outside the existing properties and logic of the existing CP-ABE access structure, we add a time trapdoor *TD* to the root node of the access policy tree as shown in Fig. 2. The time in the time trapdoor is a time point in the unified format $\mathbb{F}_t$ of the system, such as $t$ = "*year-month-day*" and $t \in \mathbb{F}_t$. A structure $\mathbb{A}$ consists of a policy tree of several nodes, and a time trapdoor *TD*. Each non-leaf node $\theta$ has two logic values $n_\theta$ denoted as the number of its child nodes and $k_\theta$ denoted as the threshold. When the user attribute set satisfies the access policy tree (in Fig. 2, the attribute set $S = \{A_0, A_1\}$ or $\{A_0, A_2\}$ satisfies the access policy) and the time trapdoor is inactive (the data has not been deleted periodically), the user can decrypt the data.

TABLE I
NOTATIONS USED IN THE PAPER

| Notation | Description |
|---|---|
| $\mathbb{A}$ | Access policy over attributes and time |
| $MSK$ | Master secret key of TA |
| $PK$ | Public parameter of the system |
| $\mathbb{F}_t$ | Unified format of time |
| $H_1$ | Hash function that maps elements in $\{0,1\}^*$ to elements in $\mathbb{G}^*$ |
| $H_2$ | Hash function that maps elements in $\mathbb{G}_T$ to elements in $\mathbb{Z}_p^*$ |
| $U_j$ | A user in the system |
| $uid_j$ | A unique identity in $\mathbb{Z}_p^*$ of user $U_j$ |
| $sk_{U_j}$ | Signature private key of user $U_j$ |
| $pk_{U_j}$ | Verification public key of user $U_j$ |
| $SK_{U_j}$ | Attribute-associated security decryption key of user $U_j$ |
| $TD$ | The time trapdoor embedded in the root node |
| $TD'$ | The time trapdoor after data deletion |
| $AAI_{index}$ | The auxiliary authentication information of MHT for the data block whose serial number is $index$ |

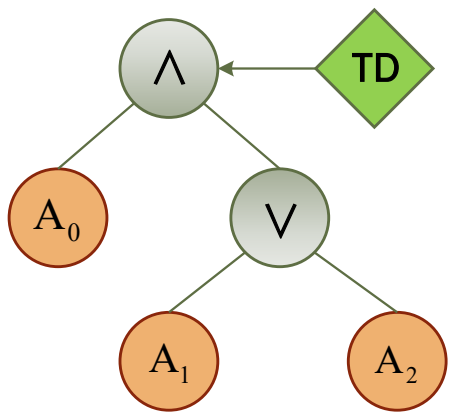

Fig. 2. Example of ATDD access policy structure.

### B. Details of Our Proposed Scheme ATDD

*1) Setup:* Firstly, TA chooses two multiplicative cyclic groups $\mathbb{G}$ and $\mathbb{G}_\mathbb{T}$ with the same prime order $p$, and the parameter $g$ is a generator of $\mathbb{G}$, the binary map $e$: $\mathbb{G} \times \mathbb{G} \to \mathbb{G}_T$ is defined on $\mathbb{G}$. Besides, TA specifies $\mathbb{F}_t$ as the time format. $H_1 : \{0,1\}^* \to \mathbb{G}^*$, $H_2 : \mathbb{G}_T^* \to \mathbb{Z}_p^*$ are both hash functions.

TA chooses $\alpha, \beta, \delta \in \mathbb{Z}_p^*$ as the master secret key randomly. The public parameter is published as

$$PK = p, \mathbb{G}, \mathbb{G}_T, g, e, H_1, H_2, \mathbb{F}_t, u = g^\beta, v = g^\delta, e(g,g)^\alpha,$$

and the master secret key is $MSK = (g^\alpha, \beta, \delta)$.

*2) Key Generation:* For each legal user $U_j$ with attribute set $S_j$, in order to verify the user's identity, TA first assigns a random $uid_j \in \mathbb{Z}_p^*$ as a unique identity for the user. Then, TA chooses a private key $sk_{U_j} \in \mathbb{Z}_p^*$, and computes its corresponding public key $pk_{U_j} = g^{sk_{U_j}}$.

For decryption, each attribute $A_i \in S_j$ is assigned a random $r_i \in \mathbb{Z}_p^*$. Then, TA computes the user's security decryption key as

$$SK_{U_j} = \Big\{ D = g^{(\alpha + uid_j)/\beta} ,$$
$$\forall A_i \in S_j : D_i = g^{uid_j} \cdot H_1(A_i)^{-r_i} , D_i' = g^{r_i} \Big\} .$$

*3) Encryption:* The data owner first encrypts the plaintext $M$ (the file name is denoted as *fname*) with the randomly selected symmetric key $\kappa \in \mathbb{G}_T$, and then encrypts the symmetric key with CP-ABE.

The node $\theta$ in the predefined access structure will be associated with a secret parameter denoted as $s_\theta$. The secret value $s_\theta$ is obtained by sharing the secret value of the parent node, and shared with the secret value of each child node (or dealt with the relevant attribute if $\theta$ is a leaf node). All parameters are calculated from the root node to the leaf node. The calculation process is as follows:

First, the data owner selects three random numbers denoted as $s_R, s_R^\tau, \hat{s}_R^\tau \in \mathbb{Z}_p$ ($s_R^\tau \neq \hat{s}_R^\tau$) for the root node $R$. $s_R^\tau$ is time trapdoor parameter and $\hat{s}_R^\tau$ is used for data deletion. The parameter $s_R'$ is calculated by $s_R' = s_R \cdot s_R^\tau$ . For the time trapdoor $TD$ related to deletion time $t \in \mathbb{F}_t$ and two secret parameters $s_R^\tau$ and $\hat{s}_R^\tau$, the data owner chooses a random parameter $r_\tau \in \mathbb{Z}_p$, and generates $TD$ as follows:

$$TD = \left(X = g^{r_\tau}, Y = s_R^\tau, Z = \hat{s}_R^\tau + H_2\left(e\left(H_1(t), v\right)^{r_\tau}\right)\right) .$$

The data owner uses $Y = \hat{s}_R^\tau$ instead of $Y = s_R^\tau$ in $TD$ to generate $TD'$ and save it locally for data deletion verification later.

For each non-leaf node $\theta$ with $s_\theta$, the data owner chooses a polynomial $f_\theta$ whose degree $d_\theta = k_\theta - 1$, and $f_\theta(0) = s_\theta$. Each of $\theta$'s child node $\vartheta$ is assigned with a unique index $ind_\vartheta$, and the parameter $s_\vartheta$ is generated by $s_\vartheta = f_\theta(ind_\vartheta)$ .

Besides, for a leaf node $\theta$ with $s_\theta$ and relevant attribute $A_\theta$, the data owner generates the child node ciphertext as follows:

$$C_\theta = g^{s_\theta}, C_\theta' = H_1(A_\theta)^{-s_\theta} .$$

Finally, the ciphertext is represented as follows:

$$CT = \left(\mathbb{A}, TD, C_0, C_1, C_2, \{C_\theta, C_\theta'\}_{\theta \in \mathbb{A}\ \&\ \theta\ is\ a\ leaf\ node}\right) ,$$

and $C_0 = E_\kappa(M)$ represents encrypting $M$ with a symmetric encryption algorithm (for example *AES*) and the secret key $\kappa \in \mathbb{G}_T^*$, $C_1 = \kappa \cdot e(g,g)^{\alpha s_R'}$, $C_2 = u^{s_R'}$.

| | | | | | | | | | |
|---|---|---|---|---|---|---|---|---|---|
| $U_j$ | *fname* | $\mathbb{A}$ | $TD$ | $C_0$ | $C_1$ | $C_2$ | $C_{\theta_1}, C'_{\theta_1}$ | $\cdots$ | $C_{\theta_l}, C'_{\theta_l}$ |

Fig. 3. Storage format of data on the cloud server.

*4) Data Upload and Timed Deletion Request:* After the data owner encrypts the data locally, the ciphertext is simply processed and uploaded to the cloud server. The data format in the cloud is shown in Fig. 3 and these data blocks (the number is $l$+7) are also MHT leaves for ensuring data deletion. After uploading the data, the user sends a *timed deletion request* (*TDR*) to the TA. The request format is as follows:

$$TDR = \Big(U_j, fname, t \in \mathbb{F}_t, index \in \mathbb{Z}_{l+7},$$
$$tag = H_1(fname||t||index), sig = tag^{sk_{U_j}}\Big) .$$

Here, the parameter *t* is the owner-defined data deletion time. The parameter *index* is the subscript of MHT leaf node which is the data block *TD* so that the cloud server can generate auxiliary verification information based on the MHT leaf node whose index value is the parameter *index*.

After receiving the *TDR*, the TA first verifies the user's identity by calculating whether the equation $e(tag, pk_{U_j}) = e(sig, g)$ is true by using user's verification public key $pk_{U_j}$.

Then, TA calculates the $tag' = H_1(fname||t||index)$ and compares it with the $tag$ to complete the integrity check.

If both of them are validated, TA stores the deletion request in a waiting list in chronological order for later use by sending a deletion time token at a specified time. Otherwise, if one of them fails, TA returns a rejection message to the user.

*5) Decryption:* After querying *CT* from the cloud server, a user $U_j$ whose attribute set is $S_j$ runs the decryption algorithm. The decryption process is from the leaf node to the root node $R$ as follows:

If the user' attribute set $S_j$ has the attribute $A_i$ which is embedded in a leaf node $\theta$, then he/she computes

$$\Gamma_\theta = \frac{e(D_i, C_\theta)}{e(D'_i, C'_\theta)} = e(g,g)^{uid_j s_\theta} \ .$$

If $A_i \notin S_j$, then $\Gamma_\theta = \perp$.

For a non-leaf node $\theta$, let $S_\theta$ be an arbitrary $k_\theta$-size set of its child nodes, and for each $\vartheta \in S_\theta$, $\Gamma \neq \perp$. If such $S_\theta$ exists, then the user figures up

$$\Gamma_\theta = \prod_{\vartheta \in S_\theta} \Gamma_\vartheta^{\prod_{\vartheta' \in S_\theta, \vartheta' \neq \vartheta} \frac{ind_{\vartheta'}}{ind_{\vartheta'} - ind_\vartheta}} = e(g,g)^{uid_j s_\theta} \ .$$

Otherwise, $\Gamma_\theta = \perp$.

For the root node $R$, if $\Gamma_\theta \neq \perp$ and the data is not periodically deleted, then the user can decrypt the ciphertext and get the plaintext $M$ by using $\Gamma_R = e(g,g)^{uid_j s_R}$. The calculation process is shown below:

$$\Gamma'_R = (\Gamma_R)^Y = e(g,g)^{uid_j s'_R},$$
$$\kappa = \frac{C_1}{e(C_2, D)/\Gamma'_R},$$
$$M = D_\kappa(C_0) = D_\kappa(E_\kappa(M)).$$

*6) Time Token Sending and Data Deletion:* When the data deletion time $t \in \mathbb{F}_t$ set by user is up, TA generates the *time token* ($TT_t$): $TT_t = H_1(t)^\delta$ .

Then, TA sends the following *deletion request* (*DR*) to the cloud server: $DR = (U_j, fname, TT_t, index)$ .

When the cloud server receives *DR*, it immediately queries the file and runs the following algorithm to overwrite old data $Y$ in *TD*: $Y = Z - H_2(e(TT_t, X)) = \hat{s}_R^\tau$ .

Then, the cloud server runs the MHT algorithm to obtain the auxiliary verification information $AAI_{index}$ of the data block $D_{index}$. Finally, the cloud server sends the *deletion proof* $DP = (U_j, fname, h_R, AAI_{index})$ to the data owner $U_j$.

*7) Verification:* After the data owner receives the *deletion proof DP*, he/she runs the same MHT algorithm with $TD'$ stored locally and $AAI_{index}$ to generate $h'_R$. If the data owner gets the result of $h_R = h'_R$, it indicates that the cloud server has indeed deleted the data.

## V. Security Analysis

*Theorem 1:* If the DBDH assumption holds, no polynomial-time adversary can selectively break ATDD with non-negligible advantage.

*Proof 1:* In DBDH game, there is an adversary $\mathcal{A}$ with a non-negligible advantage $Adv_\mathcal{A}$ in the selective security game against ATDD. We can create a simulator $\mathcal{B}$ to complete the DBDH game. The specific game process is as follows:

**Initialization**. The challenger $\mathcal{C}$ selects two cyclic groups $\mathbb{G}$ and $\mathbb{G}_T$ with prime order $p$, bilinear map $e$ and generator $g \in \mathbb{G}$. Then, $\mathcal{C}$ securely flips a random coin $b \in \{0, 1\}$. $\mathcal{C}$ sets a tuple $(A, B, C, Z) = (g^a, g^b, g^c, e(g,g)^{abc})$ if $b = 0$ or $(g^a, g^b, g^c, e(g,g)^z)$ on the contrary for random parameter $a, b, c, z \in \mathbb{Z}_p^*$. Then, $\mathcal{C}$ sends the tuple $(A, B, C, Z)$ to the simulator $\mathcal{B}$.

**Setup**. The simulator $\mathcal{B}$ randomly chooses $\alpha, \beta, \delta \in \mathbb{Z}_p^*$, and defines the time format $\mathbb{F}_t$. Then, $\mathcal{B}$ sets the hash function $H_2 : \mathbb{G}_T^* \rightarrow \mathbb{Z}_p^*$ and the hash function $H_1 : \{0,1\}^* \rightarrow \mathbb{G}^*$ which is a mapping of the form for example $H_1(A_i) = g^{m_i}$ recorded in the table. Note that the parameter $m_i$ is randomly selected from $\mathbb{Z}_p^*$. Finally, the public key $PK$ is given as

$$PK = p, \mathbb{G}, \mathbb{G}_T, g, e, H_2, \mathbb{F}_t, u = g^\beta, v = g^\delta, e(g,g)^\alpha \ .$$

$\mathcal{B}$ then sends $PK$ to the adversary $\mathcal{A}$.

**Phase 1**. $\mathcal{A}$ submits an access policy $\mathbb{A}$ and an attribute set $S' = (A_1, A_2, A_3, \cdots, A_{l_1})$ to request for a security key. Besides, all the attributes in $\mathbb{A}$ form the attribute set $S_\mathbb{A}$. $\mathcal{B}$ finds a set $\Omega$ which is the smallest complement of $S'$ in $S_\mathbb{A}$. Note that there may not be a unique $\Omega$. See the paper [12] for detailed analysis. $\mathcal{B}$ chooses the parameter $r_i$ for each attribute in $S'$ randomly, and generates the following security keys:

$$D = (C \cdot g^\alpha)^{1/\beta},$$
$$\{D_i = C \cdot H_1(A_i)^{r_i} = C \cdot (g^{d_i})^{r_i}, D'_i = g^{r_i}\}_{A_i \in S'}.$$

Then $\mathcal{B}$ sends the security keys to $\mathcal{A}$.

Next, we set up the polynomials for the access policy $\mathbb{A}$.

If $S'$ satisfies the access policy sub-tree $\mathbb{A}_\theta$, we first set a polynomial $f_\theta$ with correct degree constraints, and $f_\theta(0) = s_\theta$. Each child node $\vartheta$ obtains $s_\vartheta = f_\theta(ind_\vartheta)$ which is known to $\mathcal{B}$.

If $S'$ does not satisfy the access policy sub-tree $\mathbb{A}_\theta$, we first define a polynomial $f_\theta$ with correct degree constraints, and set $g^{f_\theta(0)} = g^{s_\theta}$. Then, for each the $\theta$'s child node $\vartheta$, if $\vartheta \in \Omega$, we calculate $g^{s_\vartheta} = g^{f_\theta(ind_\vartheta)}$ which is only known to $\mathcal{B}$. Otherwise, we choose a random $s_\vartheta \in \mathbb{Z}_p$, then fix the remaining unsatisfied points of $f_\theta$ to complete the definition of the polynomial.

Against the challenge policy $\mathbb{A}$, $\mathcal{B}$ uses $g^{f_R(0)} = A^{1/TD} = g^{a/TD}$ as input, so we can get hidden information $s_R = a/TD$. Here, $A$ is the element of the DBDH tuple and $TD$ is the time trapdoor.

**Challenge**. $\mathcal{A}$ submits two challenge messages $M_0$ and $M_1$ to $\mathcal{B}$, and $\mathcal{B}$ flips a secure coin $d \in \{0, 1\}$. For each attribute $A_i \in S_\mathbb{A}$: if $A_i \notin \Omega$, then $C_i = B^{f_i(0)}$, $C'_i = (B^{m_i})^{f_i(0)}$; otherwise, $C_i = g^{f_i(0)}$, $C'_i = (g^{m_i})^{f_i(0)}$. For the time trapdoor, $\mathcal{B}$ generates $TD = s_R^\tau$.

The ciphertext *CT* is as follows:

$$CT = \mathbb{A}, M_d \cdot \frac{e(C \cdot g^\alpha, A)}{Z}, u^s = A^\beta, \{C_i, C'_i\}_{A_i \in S_\mathbb{A}}, TD \ .$$

If $b=0$, then $Z=e(g,g)^{abc}$. We let the security key of unsatisfied attribute $A_i \in \Omega$ be $D_i = g^{bc}\cdot (g^{d_i})^{r_i}$, $D_i' = g^{r_i}$.

Suppose the Lagrange interpolation for secret $s_R$ is

$$s_R = \sum_{A_i \in \mathcal{S}} \xi_i \cdot f_i(0) \ ,$$

for any attribute set $S$ that satisfies $\mathbb{A}$. Then the secret $s$ can be calculated as $s = s_R \cdot TD$.

The $\Gamma_R$ can be calculated as

$$\begin{aligned}\Gamma_R &= \left(\prod_{A_i\in\mathcal{S}} \Gamma_i^{\xi_i}\right)^{TD} = \left(\prod_{A_i\in\mathcal{S}} \left(\frac{e\left(D_i, C_\theta\right)}{e\left(D_i', C_\theta'\right)}\right)^{\xi_i}\right)^{TD} \\ &= \left(e(g,g)^{bc}\right)^{\left(\sum_{A_i\in\mathcal{S}} \xi_i f_\theta(0)\right)\cdot TD} = e(g,g)^{abc} \ .\end{aligned}$$

Therefore, $CT$ is a valid random encryption of $M_d$.

Otherwise, if $b=1$, then $Z=e(g,g)^z$. $Z$ is randomly chosen from $\mathbb{G}_T$ in $\mathcal{A}$'s opinion, which means $CT$ has nothing about $M_d$.

**Phase 2**. Repeat Phase 1. In this phase, the action that $\mathcal{A}$ requests a security key for attribute $A_i \in S_{\mathbb{A}} - \Omega$ may lead to the abortion of the simulation and it occurs with probability $1-q_N$. This paper does not analyze the specific value of $q_N$.

**Guess**. $\mathcal{A}$ submits a guess $d'$ of $d$. If $d'=d$, $\mathcal{B}$ will output $b'=0$ to indicate that the tuple of DBDH game is a valid BDH-tuple. Instead, the tuple is random.

We assume that the distributions $b$ and $d$ are independent. Let $\mathcal{X}$ denoted the event of simulation abortion. First, we analyze the case $\mathcal{B}$ has not abort the simulation. At this point we have:

$$\begin{aligned}\Pr\left[b'=b|\bar{\mathcal{X}}\right] &= \Pr\left[b'=b|b=1,\bar{\mathcal{X}}\right]\cdot \Pr\left[b=1|\bar{\mathcal{X}}\right] \\ &\quad + \Pr\left[b'=b|b=0,\bar{\mathcal{X}}\right]\cdot \Pr\left[b=0|\bar{\mathcal{X}}\right] \\ &= 1/2\times 1/2 + 1/2\times(1/2+Adv_{\mathcal{A}}) \\ &= 1/2Adv_{\mathcal{A}} + 1/2.\end{aligned}$$

Then, we consider the case when $\mathcal{B}$ aborts the simulation, $b\prime$ is randomly chosen, and the accuracy of the guess is $\frac{1}{2}$.

Finally, the advantage of $\mathcal{B}$ in DBDH game in general is as

$$\begin{aligned}Adv_{\mathcal{B}} &= \Pr\left[b'=b|\bar{\mathcal{X}}\right]\cdot \Pr\left[\bar{\mathcal{X}}\right] \\ &\quad + \Pr\left[b'=b|\mathcal{X}\right]\cdot \Pr\left[\mathcal{X}\right] - 1/2 \\ &= (1/2Adv_{\mathcal{A}} + 1/2)\times q_N + 1/2\times(1-q_N) - 1/2 \\ &= (q_N/2)Adv_{\mathcal{A}}.\end{aligned}$$

In summary, our proposed scheme ATDD is secure against CPA for the adversary lacks adequate attribute-related keys.

## VI. Performance Analysis

In this section, we present the experimental results of our proposed scheme ATDD to demonstrate the practicality. We conduct our experiments on the Ubuntu 16.04 64-bit computer with Intel(R) Core(TM) i5-10400F CPU @ 2.90GHz. We use the Charm-Crypto 0.50 python library to implement our proposed scheme ATDD on the Visual Studio Code 1.60.1. The bilinear groups of prime order are created using super singular curve SS512 which meets the security requirements and the symmetric encryption phase is performed using AES

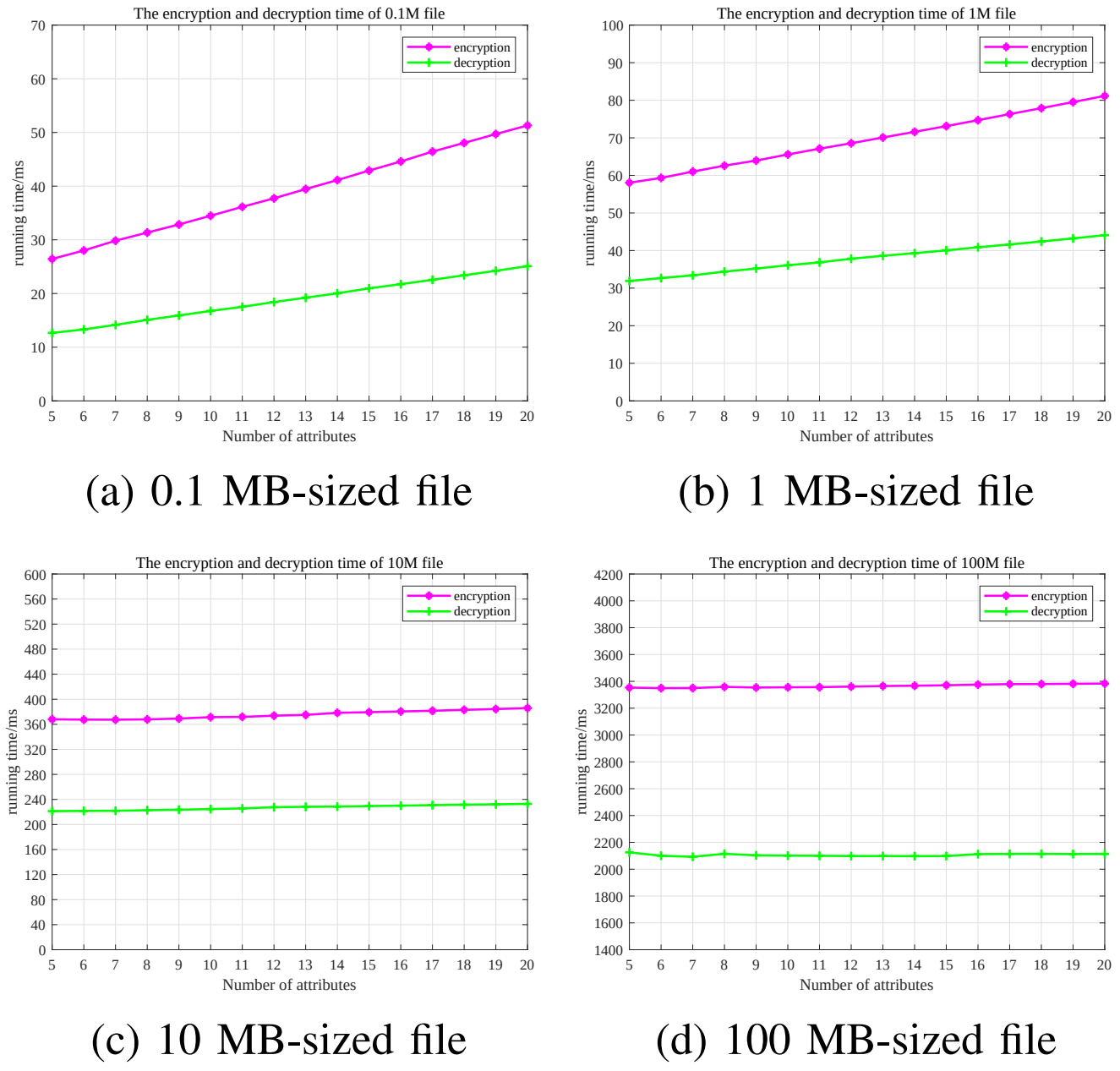

(a) 0.1 MB-sized file (b) 1 MB-sized file

(c) 10 MB-sized file (d) 100 MB-sized file

Fig. 4. The encryption time and the decryption time of four files with different sizes under the condition of different numbers of attributes.

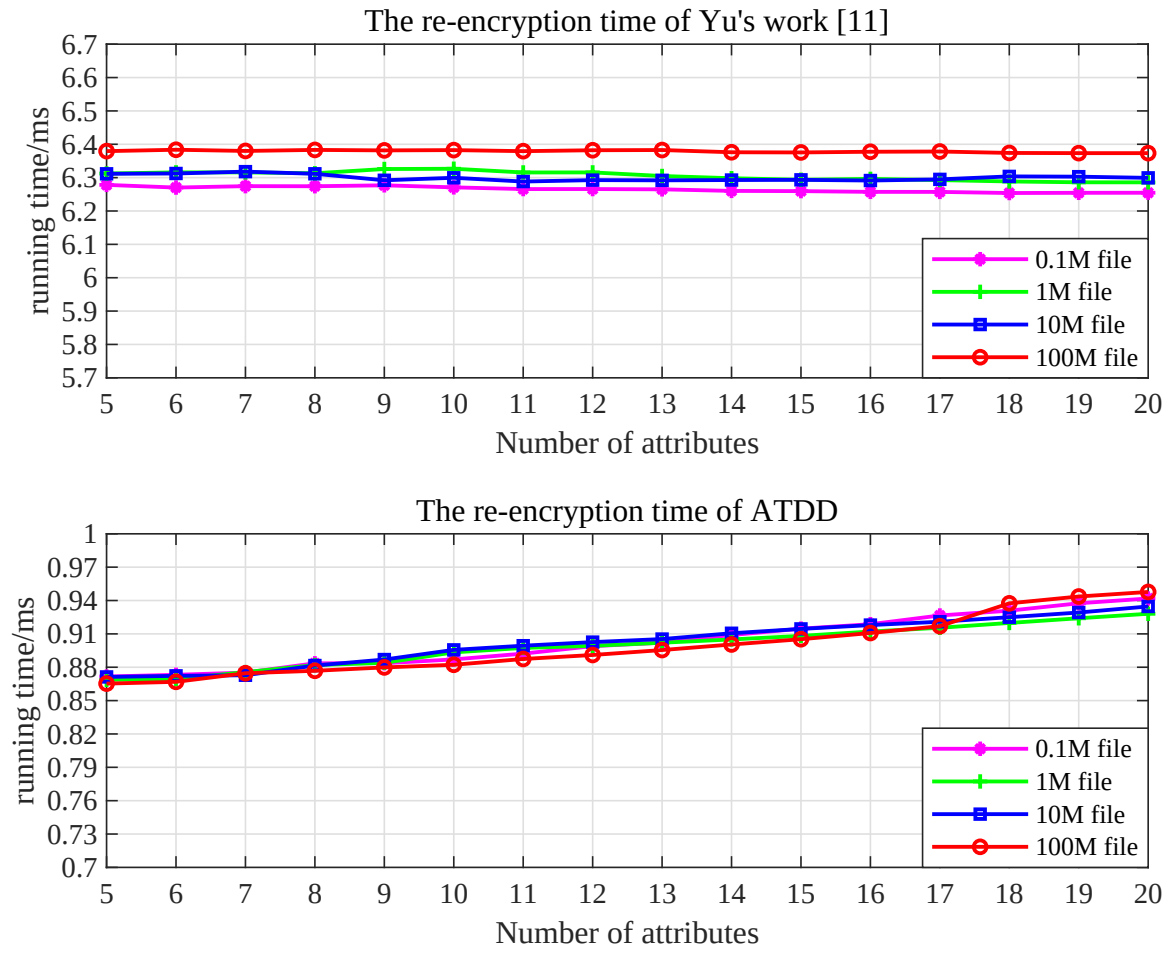

Fig. 5. The re-encryption time of four files with different sizes under the condition of different numbers of attributes (Top: Yu *et al.*'s work [11], Bottom: Our proposed scheme ATDD).

encryption. We will analyze the time cost of each system component separately.

Firstly, we test the encryption time and the decryption time of four files with different sizes (i.e., 0.1 MB, 1 MB, 10 MB, and 100 MB) when the access policy is embedded with different numbers (5-20) of attributes. The experimental results are shown in Fig. 4. The encryption time and the decryption time of files below 10 MB remains in the acceptable range of users and the obvious trend is that the encryption time and the decryption time increases as the number of attributes increases. When the file size gets larger (e.g., 100 MB), symmetric encryption and decryption consumes most time. Besides, the encryption time is below 3.6 s and decryption time is below 2.3 s, which is also acceptable to the users.

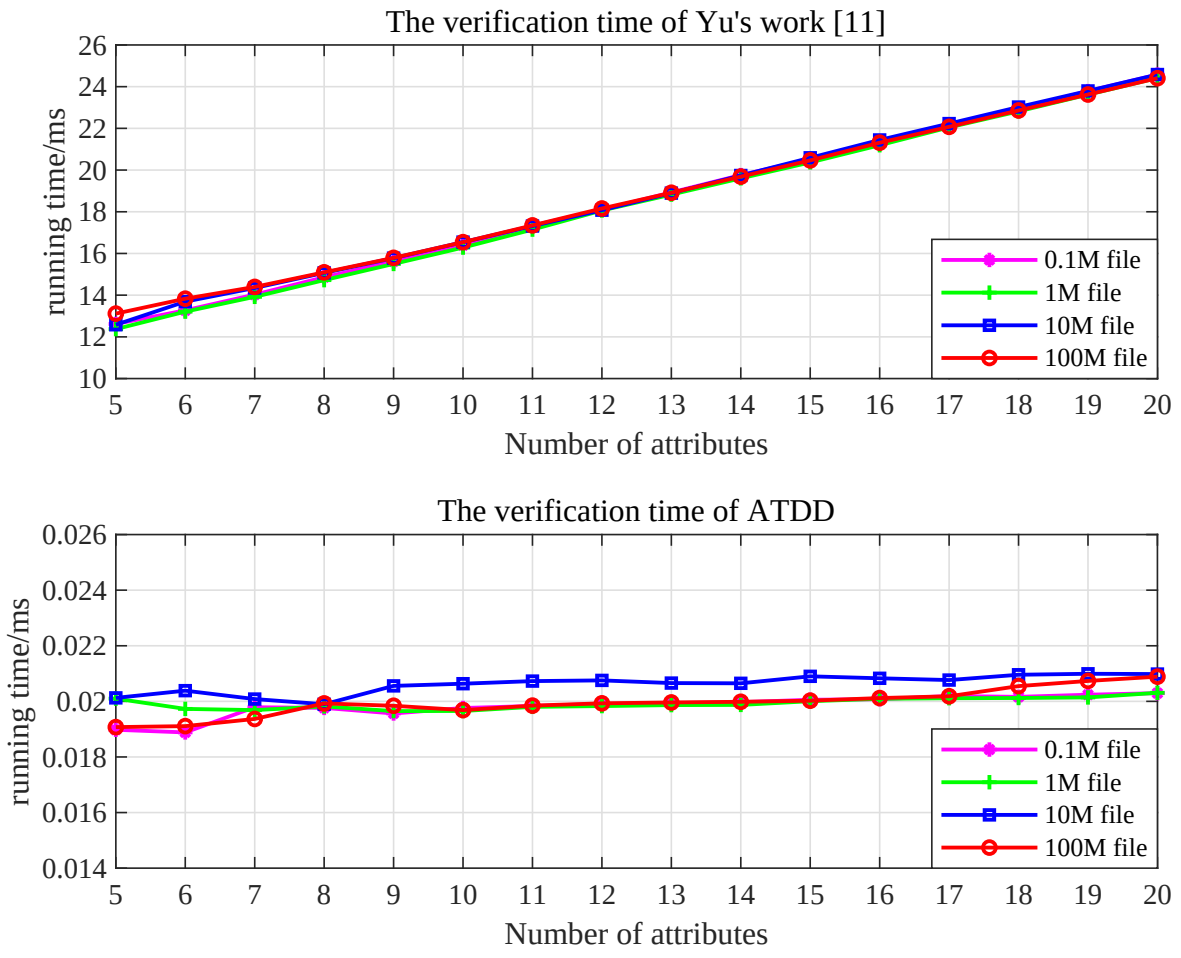

Fig. 6. The deletion verification time of four files with different sizes under the condition of different numbers of attributes (Top: Yu *et al.*'s work [11], Bottom: Our proposed scheme ATDD).

Secondly, the extra overhead of the TA is very small. It only needs two bilinear operations to verify the user's identity when the user requests timed deletion of data and one elliptic curve exponential operation to generate the deletion time token.

Thirdly, the additional overhead of the cloud server includes one-step bilinear computation and several step hashing algorithms to generate auxiliary verification information. Fig. 5 shows the time cost comparison between Yu *et al.*'s work [11] and our proposed scheme ATDD in the re-encryption process. In our ATDD, the cloud server can complete the data deletion task and generate deletion auxiliary verification information in less than 0.96 ms. In Yu *et al.*'s work, it takes about 6.3 ms to delete data because of multiple powers of large integers and elliptic curves. In contrast, our work takes less time in the data deletion phase.

Finally, we test the time cost of user deletion verification. It can be concluded from both the algorithm and the Fig. 6 that the time cost of deletion verification has nothing to do with the file size and the number of attributes. In our ATDD, it takes about 0.02 ms to complete the deletion verification process. In Yu *et al.*'s work, the deletion verification process is equivalent to the process of running the attribute-based decryption which decrypts the symmetric secret key, so the time increases with the number of attributes, and the time is more than 12 ms regardless of the file size. Hence, the cost of deletion verification process in our ATDD is very small and has little effect on the user's computation resources.

To sum up, our scheme brings about small extra computation costs and has good practical value.

## VII. Conclusion

The fine-grained assured time-sensitive data deletion (ATDD) scheme in cloud storage is proposed in this paper. The security and the practicability of our proposed scheme ATDD are demonstrated. In the future, assured cloud data deletion combining time and location factors deserves investigation.

## Acknowledgments

This work was supported by the National Key Research and Development Program of China under Grant 2018YFB0804102, the National Natural Science Foundation of China under Grant 61802357, and the Fundamental Research Funds for the Central Universities under Grant WK3480000009.